\documentclass{PoS}

\usepackage{amsmath}
\usepackage{cite}

\newcommand{\nn}{\nonumber \\}
\newcommand{\no}{\nonumber}
\newcommand{\imag}{\mathrm{Im}\,}

\newcommand{\eq}[1]{Eq.~(\ref{#1})}
\newcommand{\eqsand}[2]{Eqs.~(\ref{#1}) and (\ref{#2})}

\newcommand{\fig}[1]{Fig.~\ref{#1}}

\newcommand{\gev}{\,\mbox{GeV}}
\newcommand{\mev}{\,\mbox{MeV}}
\newcommand{\Dbar}{\,\overline{\!D}}
\newcommand{\DorDbar}{\raisebox{7.7pt}{$\scriptscriptstyle(\hspace*{8.5pt})$}
  \hspace*{-10.7pt}\!\Dbar}
\newcommand{\KorKbar}{\raisebox{7.7pt}{$\scriptscriptstyle(\hspace*{8.5pt})$}
  \hspace*{-10.7pt}\overline{K}}
\newcommand{\lqcd}{\ensuremath{\Lambda_{\rm QCD}}}
\newcommand{\dm}{\ensuremath{\Delta m}}
\newcommand{\dg}{\ensuremath{\Delta \Gamma}}
\newcommand{\epm}[2]{
  \raisebox{-0.5ex}{\shortstack[l]{$\scriptstyle+#1$\\$\scriptstyle-#2$}}}
\newcommand{\beq}[1]{\begin{align} #1 \end{align}}
\newcommand{\beqin}[1]{$#1$}

\newcommand{\ket}[1]{| #1\rangle}
\newcommand{\braket}[3]{\langle #1|#2\ket{#3}}
\newcommand{\lt}{\left}
\newcommand{\rt}{\right}

\title{Charm decays}

\ShortTitle{Charm decays}

\author{\speaker{Ulrich Nierste} \\
        Institut f\"ur Theoretische Teilchenphysik, Karlsruher
  Institut f\"ur Technologie, 76131 Karlsruhe, Germany
        \\
        E-mail: \email{ulrich.nierste@kit.edu}}

      \abstract{I discuss hadronic decays of $D$ mesons with emphasis on
        the recent discovery of charm CP violation in
        $D^0\to K^+K^-,\pi^+\pi^-$ decays. The measured difference
        $\Delta a_{CP} \, \equiv \; a_{CP}^{\mathrm{dir}}(D^0\rightarrow
        K^+K^-) - a_{CP}^{\mathrm{dir}}(D^0\rightarrow\pi^+\pi^-) = \;
        (-15.4\pm 2.9)\cdot 10^{-4}$ of two direct CP asymmetries
        exceeds the SM prediction by a factor of 7. A possible
        explanation is an enhancement of the penguin amplitude entering
        $ a_{CP}^{\mathrm{dir}}$ by QCD effects which are not understood
        yet. Alternatively, $\Delta a_{CP}$ could be dominated by
        contributions from new physics. In order to distinguish these
        two hypotheses further CP asymmetries should be measured. To
        this end CP asymmetries resulting from the interference of two
        tree-level amplitudes auch as
        $a_{CP}^{\mathrm{dir}}(D^0\rightarrow K_SK_S)$ or
        $a_{CP}^{\mathrm{dir}}(D^0 \to
        \raisebox{7.7pt}{$\scriptscriptstyle(\hspace*{8.5pt})$}
        \hspace*{-10.7pt}\overline{K}^{\;\,*0} K_S)$ are especially interesting.}

        \FullConference{18th International Conference on B-Physics at
              Frontier Machines - Beauty2019 -\\
              29 September 
             -- 4 October, 2019\\
          Ljubljana, Slovenia}

\begin{document}

\section{Overview}
The charm event of the year 2019 was the announcement of March 21,\\
\centerline{\it LHCb sees a new flavour of matter-antimatter asymmetry,}
presenting the first observation  of CP violation in charm decays. The LHCb
collaboration has measured the difference of two direct CP asymmetries
\cite{Aaij:2019kcg}:
\begin{align}
  \Delta a_{CP} \, & \equiv \;\;a_{CP}^{\mathrm{dir}}(D^0\rightarrow K^+K^-) -
  a_{CP}^{\mathrm{dir}}(D^0\rightarrow\pi^+\pi^-) \nn
   & = \;\; (-15.4\pm 2.9)\cdot 10^{-4}. \label{exp}
\end{align}

\begin{figure}[tb]
\centerline{  
\includegraphics[height=3cm]{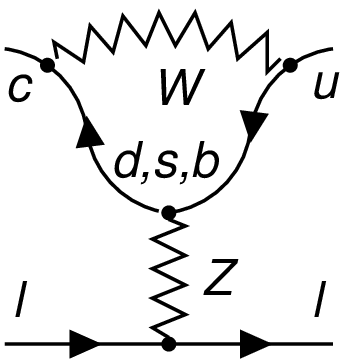}
\hspace{2cm}\includegraphics[height=3.3cm]{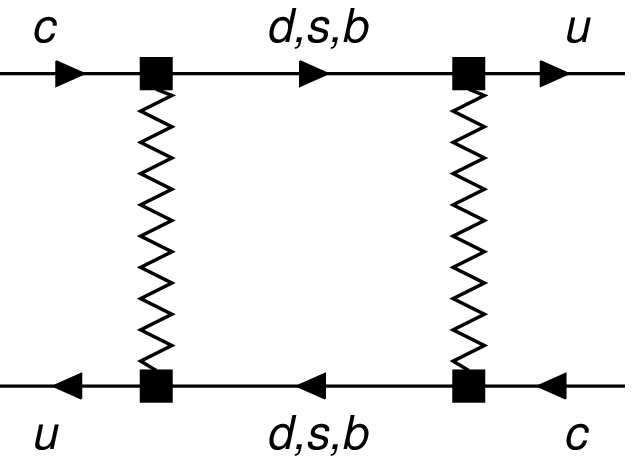}}
\caption{FCNC amplitudes: $Z$ penguin diagram contributing to $D^0\to \ell^+\ell^-$
  (left) and $D$--$\bar D$ mixing box diagram (right). \label{pengbox}}
~\\[-3mm]\hrule
\end{figure}

Before discussing the theory aspects of this measurement I give a short
overview on the role of charm decays in particle physics and the methods
and difficulties of theory predictions. While weak decays of charmed
hadrons are not useful for the metrology of the
Cabibbo-Kobayashi-Maskawa (CKM) matrix, they have a unique role in
probing new physics in the flavour sector of up-type
quarks. Flavour-changing neutral current (FCNC) amplitudes (see
\fig{pengbox} for examples) involve the CKM combinations
$\lambda_d=V_{cd}^*V_{ud}$, $\lambda_s=V_{cs}^*V_{us}$, and
$\lambda_b=V_{cb}^*V_{ub}$ associated with $d$, $s$, and $b$ quarks,
respectively, on internal lines of the FCNC loop diagrams. CKM unitarity
$\lambda_d+\lambda_s+\lambda_b=0$ allows us to eliminate one of these
CKM combinations.  If we write $p\equiv \sum_q \lambda_q p(m_q)$ for the
penguin diagram in \fig{pengbox} and choose to eliminate $\lambda_d$, we
find $p= \lambda_s [p(m_s)-p(m_d)] + \lambda_b [p(m_b)-p(m_d)]$. The
loop contribution with $\lambda_b$ is tiny because of
$|\lambda_b|\sim 10^{-4}$, while the contribution proportional to
$\lambda_s\simeq \lambda = 0.22$ vanishes in the limit $m_d=m_s$
(corresponding to unbroken U-spin symmetry) and is therefore heavily
suppressed by the Glashow-Iliopoulos-Maiani (GIM) mechanism.  The latter
feature also makes it impossible to predict FCNC processes in a reliable
way.  For example, a perturbative calculation of the loop function
$p(m_s)-p(m_d)$ involving internal $d$ and $s$ quarks in the penguin
diagram of \fig{pengbox} gives a result proportional to
\begin{align}
  \frac{G_F}{M_Z^2}\quad\cdot\quad
&\underbrace{(m_s-m_d)} \;\quad\;\cdot \quad\;\, \underbrace{(m_s+m_d)}\nn
& \!\overbrace{\parbox[t]{1.7cm}{\centering{U-spin}\\
  \centering{breaking}\\  \centering{ GIM}}}
\;\;
\overbrace{\parbox[t]{3.5cm}{\centering{ artefact of}\\ 
                            \centering{perturbation theory}}}
\end{align}
The presence of small quark masses below the QCD scale
$\lqcd\sim 400\mev$ indicates that the perturbative calculation is not
trustworthy.  While the factor $m_s-m_d$ correctly catches the linear
U-spin breaking of the amplitude, the factor $m_s+m_d$ occurs, because
the left-chiral nature of the $W$ coupling requires an even number of
left-right flips on the internal quark line. This factor is an artefact
of perturbation theory, non-perturbative QCD provides other sources of
left-right flips, for instance the quark condensate. The only
experimentally established FCNC transition in charm physics is the
$D$--$\bar D$ mixing amplitude, the mass and width difference between
the two neutral $D$ eigenstates (normalized to the total width $\Gamma$)
are (HFLAV) \cite{Amhis:2019ckw}
\begin{align}
x=\frac{\dm}{\Gamma}=0.39\epm{0.11}{0.12}\%, \qquad
  y=\frac{\dg}{2\Gamma}=0.651\epm{0.063}{0.069}\% .
  \label{xy}
\end{align}
These numbers exceed the naive perturbative result 
of the box diagram in \fig{pengbox} by far.
Still our theoretical understanding of $D$--$\bar D$ mixing is too poor
to conclude whether the measurements in \eq{xy} involve new physics
contributions or not.  Thus while charm FCNC transitions are highly
suppressed in the Standard Model (SM), our insufficient understanding of
low-energy QCD effects limits their use as new-physics analyzers.

The other avenue to new physics are measurements of CP asymmetries.
Hadronic weak decays of the $D$ mesons \beq{ D^+ \sim c \bar d,\qquad D^0
  \sim c \bar u,\qquad D_s^+ \sim c \bar s,\quad } are denoted
Cabibbo-favored (CF) or singly or doubly Cabibbo-suppressed (SCS or
DCS), if the decay amplitude is proportional to $\lambda^0$,
$\lambda^1$, or $\lambda^2$, respectively. Non-zero CP asymmetries
require the interference of two amplitudes with different CP-violating
phases, which implies that all SM predictions for charm CP asymmetries
involve the suppression factor
$\imag\frac{\lambda_b}{\lambda_s}=-6\cdot 10^{-4}$.  We may categorize
the detectable CP asymmetries by their origin from
\begin{itemize}
\item \emph{box--tree},
\item \emph{penguin--tree},  or
\item \emph{tree--tree}  
\end{itemize}
interference.
\begin{figure}[tb]
\begin{minipage}{0.65\textwidth}
  \centerline{  
  \parbox[b]{4.5cm}{\includegraphics[width=25mm,angle=270]{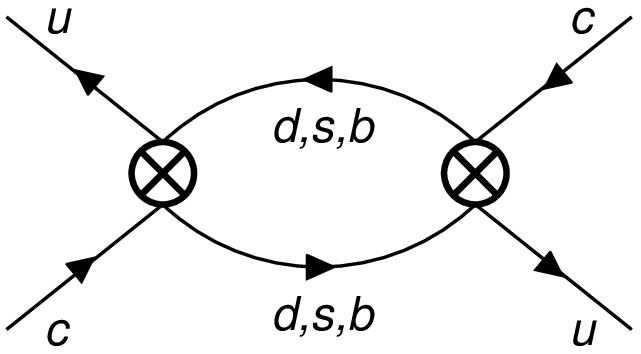}}
\hspace{1cm}\parbox[b]{3cm}{\includegraphics[width=25mm,angle=270]{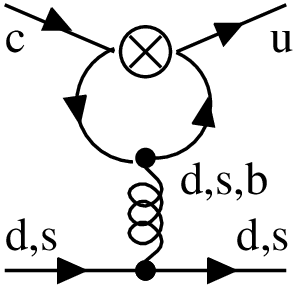}}}
\end{minipage}\hfill
\begin{minipage}{0.33\textwidth}
\centerline{
    \includegraphics[height=30mm ]{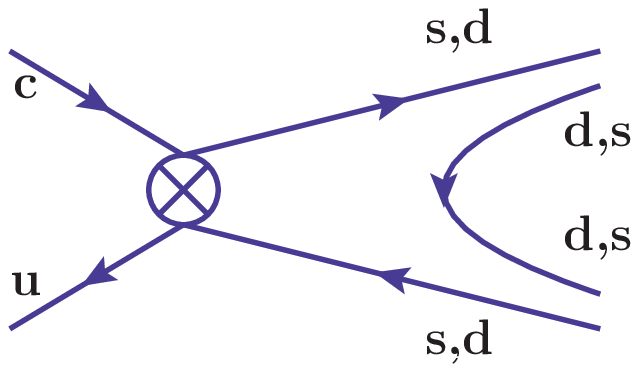}}
\end{minipage}
\caption{Box diagram describing $D$--$\bar D$ mixing, penguin diagram
  contributing to SCS decays, and tree-level exchange diagrams
  contributing to CP asymmetries in decays into two neutral kaons. The cross denotes a
  $W$ propagator contracted to a point as in the Fermi theory. \label{bpt}}
~\\[-3mm]\hrule
\end{figure}  
The first category contains mixing-induced CP asymmetries like
$a_{CP}^{\rm mix}(D^0(t)\to K^+ \pi^-)$. Most direct CP asymmetries
are in the second category and the LHCb measurement in \eq{exp} 
has established penguin-tree interference, if interpreted within the SM.
The CP asymmetries of the third category arise from the
interference of the $c\to u s\bar s$ and $c\to u d\bar d$ amplitudes.
\fig{bpt} shows sample diagrams for the three categories of CP
asymmetries.

Theoretical studies of weak decays of 
charmed hadrons heavily utilize the approximate SU(3)$_{\rm F}$ symmetry
of QCD. The QCD lagrangian is invariant under unitary rotations of the
light quark triplet $(u,d,s)$ in the limit $m_u=m_d=m_s$. The SU(2)
subgroup of unitary rotations of $(u,d)$ is the strong isospin (I-spin)
symmetry; the counterpart for $(s,d)$ is 
the above-mentioned U-spin symmetry and 
V-spin refers to rotations of $(s,u)$. The I-spin breaking of
QCD scales like $(m_d-m_u)/\lqcd \sim 0.02$, while U-spin holds to
to an accuracy of order $(m_s-m_d)/\lqcd \sim 0.3$.

\section{CP violation in penguin-tree interference}
It is helpful to use $\lambda_d+\lambda_s+\lambda_b=0$ to decompose the
amplitude $\mathcal{A}^{\mathrm{SCS}}$ of the SCS decay  of a
charged or neutral $D$ meson into two light mesons $M,M^\prime$ as
\cite{Golden:1989qx}%
\beq{\mathcal{A}^{\mathrm{SCS}}(MM^\prime) \equiv \lambda_{sd} A_{sd}
  (MM^\prime) \, - \, \frac{\lambda_b}{2} A_{b}(MM^\prime)
  \label{deco}}%
with \beq{\lambda_{sd}=\frac{\lambda_s-\lambda_d}{2}&\qquad
  \mbox{and}\quad -\frac{\lambda_b}{2}=\frac{\lambda_s+\lambda_d}{2}.}%
If we write the effective hamiltonian as%
\beq{H\,=\; \lambda_d H_d+\lambda_s H_s+\lambda_b H_b+\mbox{h.c.} }%
with %
\beq{H_q\,=\; 4 \frac{G_F}{\sqrt2}\,\lt[\, C_1\, \bar u_L^\alpha \gamma_\mu
  c_L^\alpha \, \bar q_L^\beta \gamma^\mu q_L^\beta \;+\; C_2\, \bar u_L^\alpha
  \gamma_\mu c_L^\beta\, \bar q_L^\beta \gamma^\mu q_L^\alpha \,\rt],}%
where $G_F$ is the Fermi constant, then%
\beq{ A_{sd} (MM^\prime) = \braket{MM^\prime}{H_s-H_d}{D}, \qquad A_b
  (MM^\prime) = \braket{MM^\prime}{H_s+H_d-2 H_b}{D}. \label{uspin}} %
$H_q$ contains the Wilson coefficients $C_{1,2}$ with the perturbative
QCD corrections to the $W$ exchange diagram.  $C_{1,2}$ multiply the
four-quark operators describing the $W$-mediated weak interaction (where
the Fierz relation
$\bar q_L \gamma_\mu c_L \bar u_L \gamma^\mu q_L=\bar u_L \gamma_\mu c_L
\bar q_L \gamma^\mu q_L$ is used) and $\alpha,\beta$ are color indices.

The benefit of the decomposition in \eq{deco} becomes clear from
\eq{uspin}: $A_{sd}$ is a $|\Delta U|=1$ amplitude, because $H_s-H_d$
involves $\bar s s-\bar d d$ which transforms like a U-spin
triplet. Likewise $A_b$ is a $\Delta U=0$ amplitude.  In view of
$|\lambda_b|/|\lambda_{sd}|\sim 10^{-3}$ we can work to first
non-vanishing order in $\lambda_b$ and may safely replace
$\lambda_{sd}=\lambda_s+\lambda_b/2$ by $\lambda_s$.  Data on
branching ratios can be used to determine $|A_{sd}|$ for the decay modes
of interest, but are not accurate enough to give information on $|A_b|$.

To discuss the direct CP asymmetry in some SCS decay $D\to MM^\prime$ we
need \eq{deco} for $\mathcal{A}=\mathcal{A}^{\mathrm{SCS}}(MM^\prime)$
and the analogous decomposition for the amplitude
 $\overline{\mathcal{A}}$ of the 
CP-conjugate decay
$\bar D \to \bar M \bar M^\prime$, where CP$\,\ket{D}=-\ket{\bar D}$ and
CP$\,\ket{MM^\prime}=\ket{\bar M \bar M^\prime}$:
\beq{\overline{\mathcal{A}} 
    & = -\lambda_{sd}^* A_{sd} +
    \frac{\lambda_b^*}{2} A_b.  \no}%
The SM prediction for the desired CP asymmetry reads
\begin{align}
a_{CP}^{\mathrm{dir}} &\, \equiv \;\frac{
        \vert \mathcal{A}\vert^2 - \vert \overline{\mathcal{A}} \vert^2
       }{
        \vert \mathcal{A}\vert^2 + \vert \overline{\mathcal{A}} \vert^2
         } 
\; = \; \imag \frac{\lambda_b}{\lambda_{sd}}
   \; 
  \imag \frac{A_b}{A_{sd}} \, =\,  -6\cdot 10^{-4} \; \mathrm{Im}\frac{A_b}{A_{sd}}.
\label{dcp}
\end{align}  

\begin{figure}[tb]
  \centerline{  \includegraphics[width=0.23\textwidth]{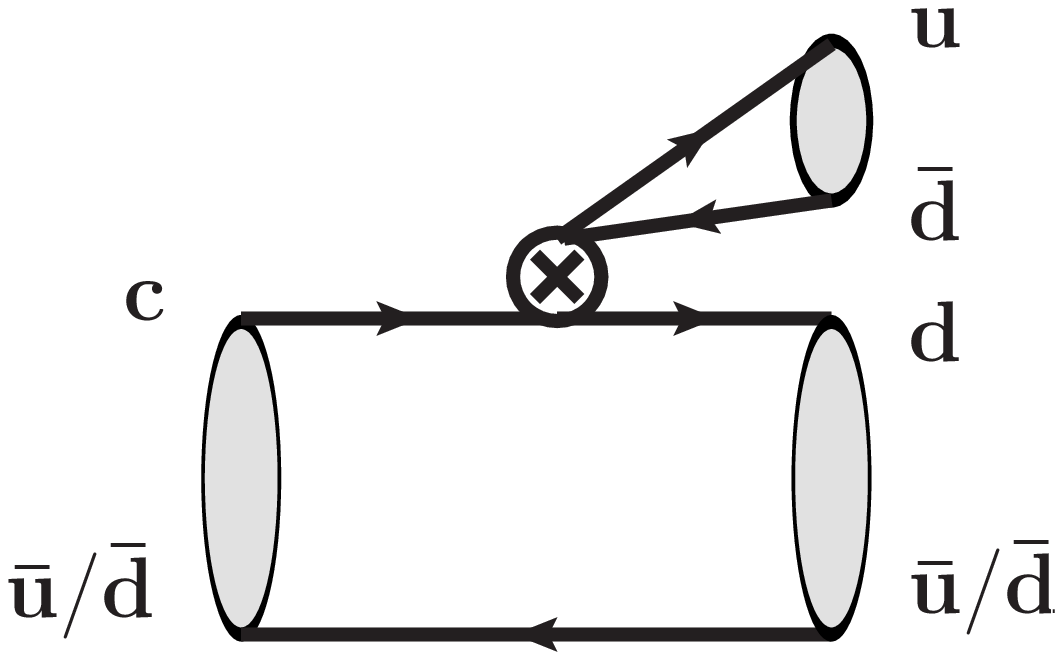}~~
\includegraphics[width=0.23\textwidth]{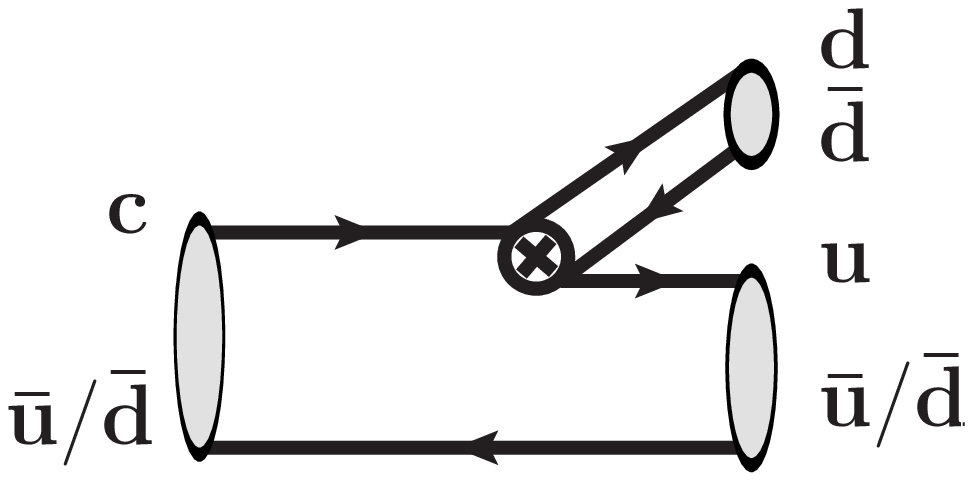}~~
\includegraphics[width=0.23\textwidth]{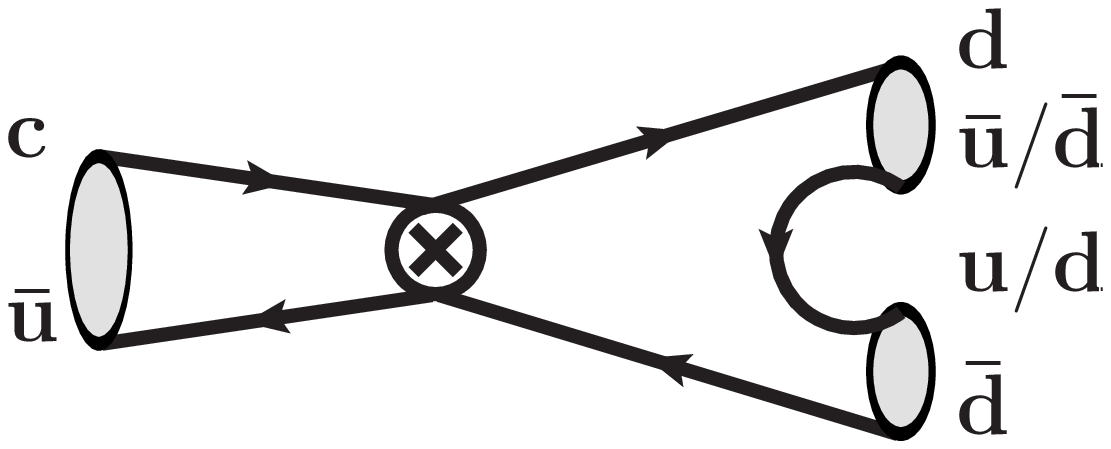}~~
\includegraphics[width=0.23\textwidth]{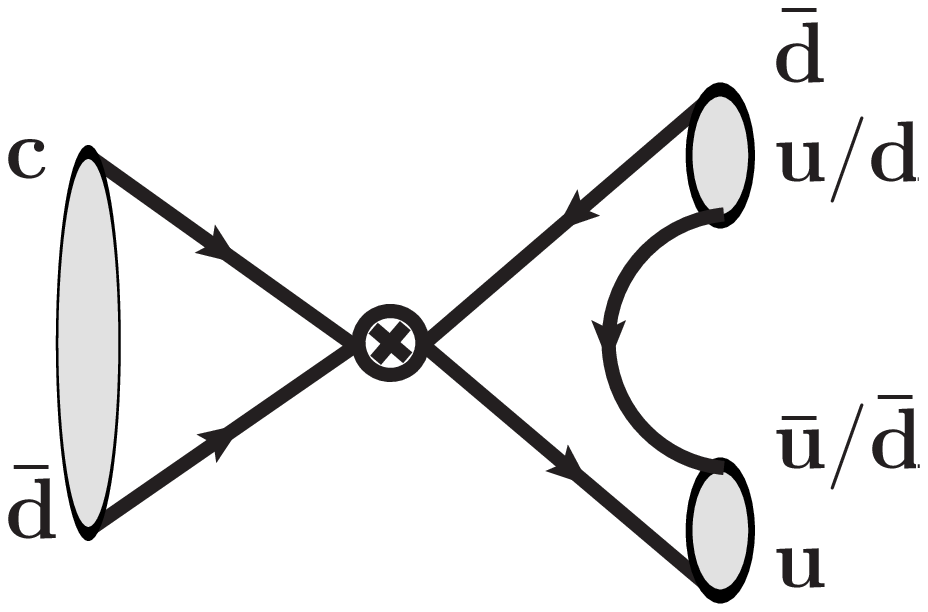}}
    \caption{Topological amplitudes \emph{color-favored tree}\ $T$,
      \emph{color-suppressed tree}\ $C$, \emph{exchange}\ $E$, 
    and \emph{annihilation}\ $A$. They are understood to include
    all perturbative and non-perturbative strong-interaction effects,
    i.e.\ one may view the diagrams as dressed with an arbitrary number
    of gluons.
    $T$, $C$, $E$, and $A$  are complex
    numbers which are determined from a global fit to data. 
    \label{topa}. }
~\\[-3mm]\hrule
\end{figure}

One can conveniently describe $\bar D \to \bar M \bar M^\prime$ decays
in terms of topological amplitudes \cite{Wang:1980ac,Zeppenfeld:1980ex}.
In the SU(3)$_{\rm F}$ limit we can express $A_{sd}$ of all
$D^0,D^+,D_s^+ \to M M^\prime$ decays for all combinations
$ M, M^\prime= \pi^{\pm,0}, K^{\pm,0}$ as linear combinations of the four
tree diagrams $T$, $C$, $E$, and $A$ shown in \fig{topa}.
Linear (i.e.\ first-order) SU(3)$_{\rm F}$ breaking can be included in
the method in a straightforward way \cite{Gronau:1995hm} . The
topological-amplitude method is mathematically equivalent
\cite{Muller:2015lua} to the decomposition of the decay amplitudes in
terms of matrix elements classified by their SU(3)$_{\rm F}$ symmetry
properties \cite{Grossman:2012ry,Hiller:2012xm}.  A global fit of all branching ratios
to the four SU(3)$_{\rm F}$ limit amplitudes of \fig{topa} returns a
poor fit. If one includes the topological amplitudes parametrising
linear SU(3)$_{\rm F}$ breaking the fit is underconstrained and one
obtains a perfect fit on a large submanifold of the parameter space
\cite{Muller:2015lua}.  By assuming upper bounds on the sizes of the
SU(3)$_{\rm F}$-breaking topological amplitudes (limiting their magnitudes
to e.g.\ 30\% of the leading $T$ amplitude) one can nevertheless derive
useful constraints on $T$, $C$, $E$, $A$ and the
SU(3)$_{\rm F}$-breaking amplitudes \cite{Muller:2015lua}.

The information from branching ratios is not sufficient to predict CP
asymmetries: $A_b$ in \eq{dcp} involves new topological amplitudes
in the  SU(3)$_{\rm F}$ limit, which cannot be constrained from
branching fractions. These are the \emph{penguin}\ amplitude $P$ and
the \emph{penguin annihilation}\ amplitude $PA$. Consider

\begin{figure}[tb]
  \centerline{  \includegraphics[width=0.43\textwidth]{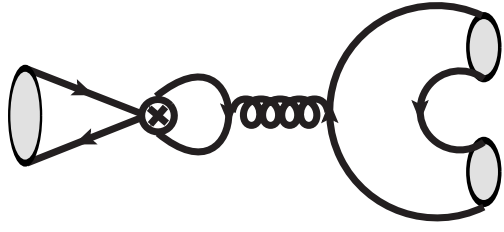}}
  \caption{Topological \emph{penguin annihilation}\ amplitude $PA_q$,
    where $q$ is the quark flavor of the loop. The perturbative gluon
    provides the hard momentum transfer from the loop to the final
    state, further soft QCD interaction is needed to arrange the correct
    color quantum numbers.\label{pa} }
~\\[-3mm]\hrule
\end{figure}

\parbox[b]{1cm}{\beqin{P_d\equiv}\\[2mm]} \includegraphics[width=0.23\textwidth]{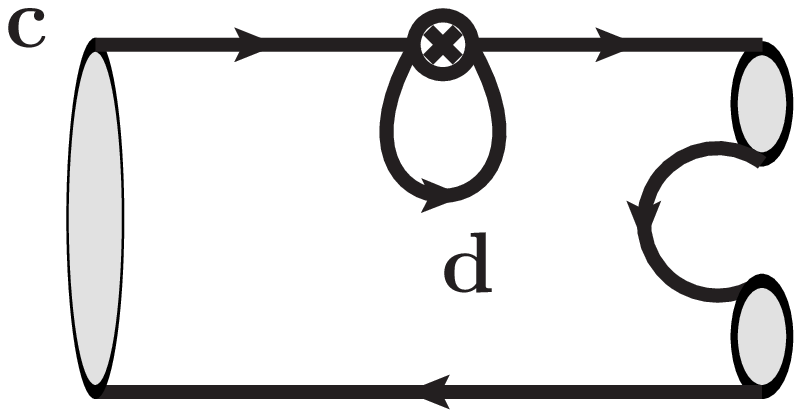}\hspace{2cm}
\parbox[b]{1cm}{\beqin{P_s\equiv}\\[2mm]} \includegraphics[width=0.23\textwidth]{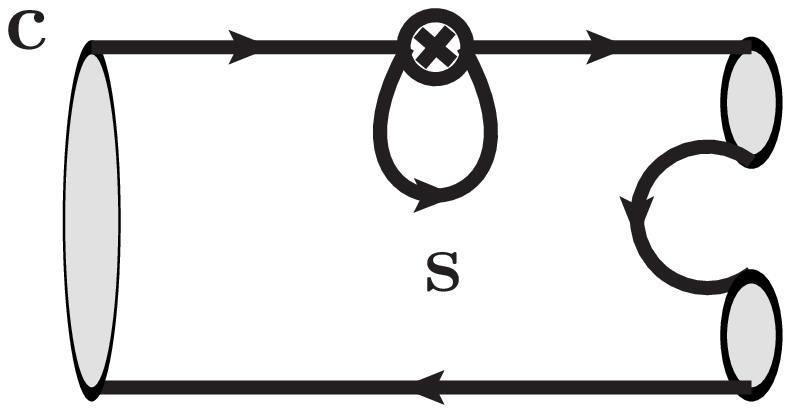}

and  the analogously defined  \beqin{P_b}. The  amplitude \beqin{A_b}
of a SCS decay involves%
\beq{P\equiv P_d+P_s-2 P_b}%
and/or the analogous combination \beqin{PA \equiv PA_d+PA_s-2 PA_b}
defined in terms of the PA amplitude in \fig{pa}.
$P$ and $PA$ are $\Delta U=0$ amplitudes and therefore do not appear in
$A_{sd}$ constrained from branching ratio data. 

In the SU(3)$_{\rm F}$ limit one finds
\beqin{A_b(\pi^+\pi^-)=A_b(K^+K^-)},
\beqin{A_{sd}(\pi^+\pi^-)=-A_{sd}(K^+K^-)} \cite{Grossman:2006jg},
and%
\beq{\imag\frac{A_b(\pi^+\pi^-)}{A_{sd}(\pi^+\pi^-)} = -
        \imag\frac{A_b(K^+K^-)}{A_{sd}(K^+K^-)} = 
        \imag \frac{P+PA}{A_{sd}(\pi^+\pi^-)} \label{su3}.}%
In the last equation $A_b(K^+K^-)=A_{sd}(K^+K^-)+P+PA$
\cite{Muller:2015rna} has been used.

A consequence of the SU(3)$_{\rm F}$ relation in \eq{su3} for
$\Delta a_{CP}$ in \eq{exp} is%
\beq{\Delta a_{CP} \stackrel{\scalebox{0.7}{\mbox{ SU(3) limit}}}{=} 2
  a_{CP}^{\mathrm{dir}}(D^0\rightarrow K^+K^-).}  Thus in the SM we
expect $\Delta a_{CP}$ to be twice as large as the individual CP
asymmetries, up to corrections from SU(3)$_{\rm F}$ breaking.
$a_{CP}^{\mathrm{dir}}(D^0\rightarrow K^+K^-) = -
  a_{CP}^{\mathrm{dir}}(D^0\rightarrow\pi^+\pi^-)$ is an example of an
SU(3)$_{\rm F}$ sum rule relating different CP asymmetries to each
other \cite{Grossman:2012ry}. One can improve such sum rules by  
including first-order breaking SU(3)$_{\rm F}$ breaking in $A_{sd}$
and e.g.\ find a refined sum rule involving the direct CP asymmetries
in $D^0\to K^+K^-$, $D^0\to \pi^+\pi^-$, and $D^0\to \pi^0\pi^0$
\cite{Muller:2015lua}. This is possible, because the global fit on $D$
branching ratios returns information on magnitudes and phases of the
topological amplitudes contributing to $A_{sd}$ for the three amplitudes.

The history of measurements of $\Delta a_{CP} $ prior to the 2019 discovery 
is as follows:\\[1mm]
\centerline{\parbox[t]{0.4\textwidth}{Previous LHCb measurements:\\
\begin{tabular}{ll}
  {2011 \cite{Aaij:2011in}:}~~&\beqin{\Delta a_{CP}=(-82\pm 21\pm 11)\cdot 10^{-4}}\\
  {2014 \cite{Aaij:2014gsa}:}~~&\beqin{\Delta a_{CP}=(-14\pm 16\pm 8)\cdot 10^{-4} } \\
  {2016 \cite{Aaij:2016cfh}:}~~&\beqin{\Delta a_{CP}=(-10\pm 8\pm 3)\cdot 10^{-4} } 
\end{tabular}}
\hspace{2cm}
\parbox[t]{0.4\textwidth}{Previous world averages {(HFLAV)}:\\
\begin{tabular}{ll}
{2015:}~~&\beqin{\Delta a_{CP}=(-25.3\pm 10.4)\cdot 10^{-4} } \\
  {2016:}~~&\beqin{\Delta a_{CP}=(-13.4\pm 7.0)\cdot 10^{-4} }
\end{tabular}}}
~\\[-1mm]
Theoretical analyses of CP asymmetries based on  SU(3)$_{\rm F}$
symmetry can relate different CP asymmetries but cannot predict
the overall size because of the a priori unknown $P$ and $PA$
amplitudes. In 2011 LHCb presented the first evidence for a non-zero
$\Delta a_{CP}$ with the value quoted above \cite{Aaij:2011in}, which
was unexpectedly large. All
SU(3)$_{\rm F}$ papers written afterwards (such as
Ref.~\cite{Muller:2015rna}) present ranges for  $\Delta a_{CP}$
compatible with the value in \eq{exp}, because they use the 2011 value
as input. This feature merely reflects the fact that  $\Delta a_{CP}$
in \eq{exp} complies with the earlier measurement presented in 
Ref.~\cite{Aaij:2011in}. 

Confronting \beq{a_{CP}^{\mathrm{dir}}  (D^0\to K^+K^-) 
\simeq \frac12 \Delta a_{CP} 
 = \frac12 (-15.4\pm 2.9)\cdot 10^{-4}} 
with \eq{dcp}  
one can solve for the imaginary part of the
{``penguin-to-tree ratio''}:
\beq{\frac12 \, \frac{A_b(K^+K^-)}{A_{sd}(K^+K^-)} \approx
  \frac{P_d}{A_{sd}(K^+K^-)} \label{pt} }
to find \cite{Grossman:2019xcj}
\beq{ \frac12 \, \frac{A_b(K^+K^-)}{A_{sd}(K^+K^-)} = 0.65\pm 0.12 .} 
Methods employing a perturbative calculation of the penguin diagram
in \fig{pengbox} give much smaller values for the ratio in \eq{pt}.
The authors of Ref.~\cite{Grossman:2019xcj} conclude that there is
either a non-perturbative enhancement mechanism of the $\Delta U=0$
amplitude $A_b$ (i.e.\ an enhancement of $P+PA$) \cite{Golden:1989qx}
or physics beyond the SM (BSM).

The momentum flowing through the penguin loop in $P$ and $PA$ are of
order 1 \gev or larger, therefore a perturbative calculation of this
loop is not unreasonable. In QCD sum rule calculations this loop is
indeed calculated perturbatively and Ref.~\cite{Khodjamirian:2017zdu} finds 
\beq{|\Delta a_{CP}| \leq   (2.0\pm 0.3)\cdot 10^{-4}, \label{kp}}
which is smaller than the experimental value by a factor of 7!
QCD sum rules are a well established method successfully describing
many quantities in $B$ physics while poorly tested in $D$ physics.
An essential ingredient of QCD sum rule calculations is the assumptions
that the combined effect of all highly excited hadronic resonances and
multi-hadron states is correctly described by a perturbative
calculation.

Next I argue that one arrives at an estimate in the ballpark of \eq{kp}
even without invoking a perturbative treatment of the penguin loop. We
need \beqin{\mathrm{Im}\frac{A_b}{A_{sd}}=
  \frac{\mathrm{Im}\,A_bA_{sd}^*}{|A_{sd}|^2}} and the numerator
$\mathrm{Im}\,A_bA_{sd}^*$ is the absorptive part of the penguin-tree
interference
term:\\
\centerline{
  \includegraphics[height=0.25\textwidth]{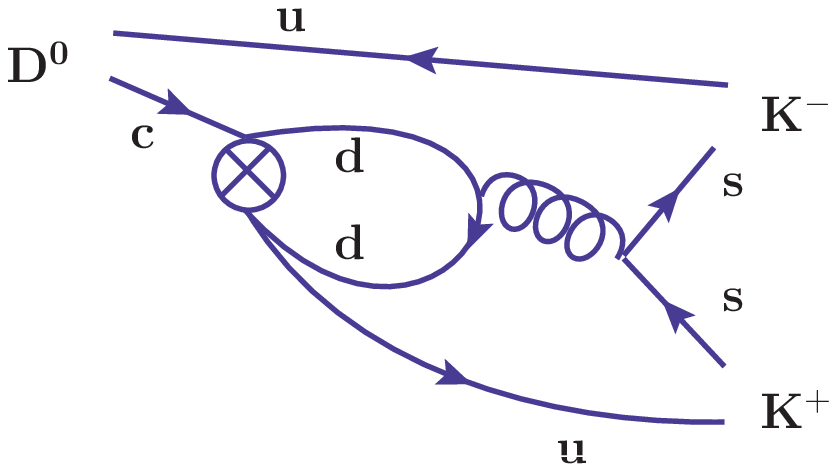}\hspace{2mm}
  \includegraphics[height=0.25\textwidth]{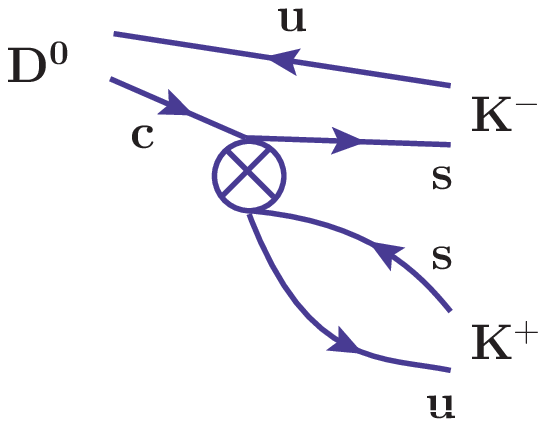}} By the optical
theorem this absorptive part is related to a \beqin{c\to u d\bar d}
decay followed by \beqin{d\bar d \to s \bar s} rescattering. This
rescattering is essential for a non-zero direct CP asymmetry and we may
discuss it without referring to the perturbative picture of quarks and
gluons.  One  contribution is \beqin{D^0\to \pi^+\pi^- \to K^+K^-}
rescattering. Each such contribution to
\beqin{\mathrm{Im}\frac{A_b}{A_{sd}}} is color-suppressed \beqin{\propto
  1/N_c} and further suppressed by a factor of \beqin{\sim 1/\pi} from
the phase space integral of the rescattering process. We conclude that
we need an enhancement factor \beqin{X} for the $\Delta U=0$ transitions
feeding $A_b$ such that \beqin{X \cdot \frac{1}{N_c\pi} \stackrel{!}{=}
  0.65\pm 0.12}. This means \beqin{X\sim 6}, thus the QCD sum rule
result of Ref.~\cite{Khodjamirian:2017zdu} has the expected size and is
not unnaturally small.  A resonant enhancement involving only the
$\Delta U=0$ channel leaves $A_{sd}$ unchanged and can therefore
accomodate $\Delta a_{CP}$ in \eq{exp} without violating data on
branching fractions which comply with the SM \cite{Muller:2015lua}.  In
Ref.~\cite{Soni:2019xko} it has been suggested that the $f_0(1710)$
resonance (having a mass close to the $D^0$ mass) could provide such an
enhancement mechanism through $D^0\to f_0(1710) \to K^+K^-$ or
$ \pi^+\pi^-$. For this mechanism to work the overlap of the $ f_0(1710)$ state
with the $K^+K^-$ or $\pi^+\pi^-$ state must be sufficiently large,
in contradiction with the expectation that a high resonance will
dominantly decay to high-multiplicity states. More insight will be
gained from measurements of the branching fractions of $ f_0(1710)$
into $K^+K^-$ or $\pi^+\pi^-$. Since in $D^0\to f_0(1710) \to
M M^\prime$ decays the final state carries the quantum numbers of the
$ f_0(1710)$ one can find SU(3)$_{\rm F}$ relations among different
CP asymmetries which are specific to this mechanism and may serve to
falsify the $ f_0(1710)$ resonance hypothesis \cite{Soni:2019xko}.

Physics beyond the SM may well affect $\Delta a_{CP}$. If the BSM
contribution to \beqin{c\to u d\bar d} or \beqin{c\to u s\bar s} comes
with an arbitrary ${\cal O} (1)$ CP phase, the suppression factor
$\imag \frac{\lambda_b}{\lambda_{sd}} \, =\, -6\cdot 10^{-4}$ is absent
and the exchange of a virtual multi-TeV particle could induce a
$\Delta a_{CP}$ in the range of \eq{exp}. Various BSM scenarios with
heavy particles are discussed in Ref.~\cite{Dery:2019ysp}. Also light
BSM particles with feeble couplings may explain the measured
$\Delta a_{CP}$; Ref.~\cite{Chala:2019fdb} studies a model with a
$Z^\prime$ boson. If the new physics couples differently to \beqin{s} and \beqin{d} quarks
(i.e.\ if it violates {U-spin} symmetry), then 
\beqin{a_{CP}^{\mathrm{dir}}(K^+K^-) \approx - a_{CP}^{\mathrm{dir}}(\pi^+\pi^-)}  
does not hold. Thus such new-physics scenarios can be distinguished from
the hypothesis of QCD enhanced $A_b$ amplitudes. To this end one must
measure one of the individual CP asymmetries
\beqin{a_{CP}^{\mathrm{dir}}(K^+K^-) } and \beqin{a_{CP}^{\mathrm{dir}}(\pi^+\pi^-)}  
or their sum. 

\section{CP violation in tree-tree interference}
Whenever the tree-level transitions \beqin{c\to u \bar d d} and
\beqin{c\to u \bar s s} interfere, the decay can have a non-vanishing
direct CP asymmetry proportional to%
\beq{\textstyle\imag \frac{V_{ud}V_{cd}^*}{V_{us}V_{cs}^*}\,=\, \imag
  \frac{-V_{us}V_{cs}^*-V_{ub} V_{cb}^*}{V_{us}V_{cs}^*}\,=\, - \imag
  \frac{V_{ub}V_{cb}^*}{V_{us}V_{cs}^*} \,\simeq\, -\imag
  \frac{\lambda_b}{\lambda_{sd}}\,\simeq\, 6\cdot 10^{-4}.}%
Tree-tree interference occurs for final states containing an
\beqin{\eta^{(\prime)}, \omega,\ldots,} or a pair of neutral Kaons like
\beqin{K_SK_S,K_SK^{*0},\ldots,} or for multibody final states  like
\beqin{K^+K^-\pi^+\pi^-} containing
all four \beqin{s,\bar s, d,\bar d} quarks. 
The topological amplitudes $E$ (in \fig{bpt} on the right) and $PA$
 (in \fig{pa}) constitute $A_b$ entering the CP asymmetry in
$D^0$ decays into two neutral Kaons. The global fit to two-body
$D^0,D^+,D_s^+$ decays into two pseudoscalars in
Ref.~\cite{Muller:2015lua} has returned a large value of $E$, so that
$A_b(K_SK_S)$ and \beqin{a_{CP}^{\mathrm{dir}}(K_S K_S)} in the SM can
be large \cite{Nierste:2015zra}: %
\beq{\vert a_{CP}^{\mathrm{dir}}(D^0\rightarrow K_SK_S)\vert \leq 1.1\%
  \qquad \mbox{@95\% C.L.} \label{acpks}}%
Throughout this talk it is assumed that the CP violation in Kaon mixing
is properly subtracted from the measured $a_{CP}^{\mathrm{dir}}$
\cite{Grossman:2011zk}.  The ratio $A_b(K_SK_S)/A_{sd}(K_SK_S)$ is
large, because $A_{sd}(K_SK_S)$ vanishes in the SU(3)$_{\rm F}$ limit,
while $A_b(K_SK_S)$ does not. The size of the $D^0\to K_SK_S$ branching
fraction (proportional to $|A_{sd}|^2$) measures the size of
SU(3)$_{\rm F}$ breaking in $E$
\cite{Muller:2015lua,Nierste:2015zra}. The maximal value in \eq{acpks}
corresponds to the maximal value of $|2 E+PA|$ returned by the fit of
Ref.~\cite{Muller:2015lua} in addition to a favorable strong phase
difference $\arg(A_b/A_{sd})=\pm \pi/2$.  More likely values for
$|a_{CP}^{\mathrm{dir}}(D^0\rightarrow K_SK_S)|$ are three times smaller
than the upper bound in \eq{acpks}. If the strong phase
$\arg(A_b/A_{sd})$ is close to zero,
$|a_{CP}^{\mathrm{dir}}(D^0\rightarrow K_SK_S)|$ will be too small to be
measured.  However, in this case one will find instead a larger
mixing-induced CP asymmetry in $D^0(t)\rightarrow K_SK_S$
\cite{Nierste:2015zra}.
     
Other interesting decay modes to study CP violation from tree--tree
interference are $D^0 \to \bar{K}{}^{*0} K_S $ and
$D^0 \to K{}^{*0} K_S $. Since the final state is not a CP eigenstate,
these decay modes offer more possibilities for CP studies.
As a special feature of these modes the CP asymmetry persists even in
the untagged sample of $\DorDbar \to  \bar{K}{}^{*0} K_S$ and one can
determine a non-vanishing CP asymmetry by just counting
$ \DorDbar \to \bar{K}{}^{*0} K_S $ and  
$ \DorDbar \to K{}^{*0} K_S $ events  \cite{Nierste:2017cua} in a sample
with equal number of $D^0$ and $\bar D{}^0$ decays. 
In real life, however, one must study the four Dalitz plots of
$D^0,\bar D{}^0 \to \,(K^-\pi^+)_{\bar{K}{}^{*0}}\,K_S$ 
and $D^0,\bar D{}^0 \to \,(K^+\pi^-)_{K{}^{*0}}\, K_S$ to
take care of interferences with other decay modes leading to
a $K^\mp \pi^\pm K_S$ final state.  

The SM prediction is \cite{Nierste:2017cua}%
\beq{|a_{CP}^{\mathrm{dir}} (D^0\to\bar{K}{}^{*0} K_S)| \leq
  0.003 \label{kstk},} %
and the same bound applies to
$|a_{CP}^{\mathrm{dir}} (D^0\to K^{*0} K_S)|$. In the
SU(3)$_{\rm F}$ limit
$a_{CP}^{\mathrm{dir}} (D^0\to\bar{K}{}^{*0} K_S) = -
a_{CP}^{\mathrm{dir}} (D^0\to K {}^{*0} K_S) $ holds. The value in
\eq{kstk} is smaller than the one in \eq{acpks}, because
$A_{sd}(\bar{K}{}^{*0} K_S)$ and $A_{sd}(K^{*0} K_S)$ do not vanish in
the SU(3)$_{\rm F}$ limit. The prediction in \eq{kstk} uses data from an
LHCb analysis of the $D^0 \to \, K^\mp\pi^\pm K_S$ Dalitz plot
\cite{Aaij:2015lsa}.

The original motivation to study CP violation in tree-tree interference
was the possibility of large CP asymmetries in the SM, i.e.\ the
$D^0\rightarrow K_SK_S$ and $D^0 \to \KorKbar{}^{\;*0} K_S $ modes were
proposed as discovery channels for CP violation in charm decays
\cite{Nierste:2015zra,Nierste:2017cua}.  Now, in view of the
experimental result in \eq{exp} the measurement of CP asymmetries from
tree-tree interference will instead give valuable insight into the
mechanism underlying the large value in \eq{exp}. For example, QCD
dynamics enhancing $P$ and $PA$ by a factor of $7$ cannot enhance
$|a_{CP}^{\mathrm{dir}}(D^0\rightarrow K_SK_S)|$ or
$|a_{CP}^{\mathrm{dir}}(D^0\rightarrow K_SK_S)|$ and
$|a_{CP}^{\mathrm{dir}}(D^0 \to \KorKbar{}^{\;*0} K_S)| $ over the
results in \eqsand{acpks}{kstk} by the same factor of $7$.  In Sec.~V of
Ref.~\cite{Nierste:2015zra} the correlation of the imprints of new
physics on various CP asymmetries is discussed.

\section{Summary}
All CP asymmetries in the SM are proportional to the small factor
$\imag \frac{\lambda_b}{\lambda_{sd}}\,\simeq\, -6\cdot 10^{-4}$, which
makes these asymmetries sensitive to new physics. The measured value in
\eq{exp} exceeds the theory prediction \cite{Khodjamirian:2017zdu} by a
factor of 7. An explanation within the SM calls for enhanced QCD effects
in $\Delta U=0$ transitions \cite{Golden:1989qx,Grossman:2019xcj} whose
origin is currently not understood.  With more precise data on other
charm CP asymmetries we can hope to find out whether a QCD effect or BSM
physics is behind $\Delta a_{CP}$ in \eq{exp} \cite{Grossman:2006jg,
  Muller:2015rna,Nierste:2015zra,Nierste:2017cua,Dery:2019ysp,Chala:2019fdb}.
This discrimination will be straightforward, if the new physics couples
differently to $d$ and $s$ quarks, so that the SU(3)$_{\rm F}$ sum rules
of Refs.~\cite{Grossman:2012ry,Muller:2015rna} are violated beyond the
expected level of SU(3)$_{\rm F}$ breaking.

\end{document}